\documentclass[12pt]{article}
\usepackage{amsmath}
\usepackage{graphicx}
\usepackage{enumerate}
\usepackage{natbib}
\usepackage{url} 

\usepackage{titling}
\usepackage{amsfonts,amsmath,amssymb}
\usepackage{geometry}

\usepackage[percent]{overpic}
\usepackage{float}
\usepackage{graphicx}
\usepackage{multirow}
\usepackage[dvipsnames, usenames]{color}
\usepackage{tabularx}
\usepackage[format=hang, indention=-1.7cm, font=small, textfont=it, labelfont=bf]{caption}
\usepackage{setspace}

\def\dsp{\def\baselinestretch{1.3}\large\normalsize}
\dsp

\usepackage{algorithm, algpseudocode}

\makeatletter
\newcounter{phase}[algorithm]
\newlength{\phaserulewidth}

\makeatother

\usepackage{amsthm}

\theoremstyle{definition}

\usepackage{bm}
\usepackage[utf8]{inputenc}
\usepackage{amssymb}
\usepackage[justification=centering]{caption}

\usepackage{graphicx}
\usepackage{mathtools}
\usepackage{booktabs}
\usepackage{amstext} 
\usepackage{epstopdf}
\usepackage{array} 
\usepackage{booktabs}
\usepackage{caption}
\usepackage{subcaption}

\usepackage{natbib}

\usepackage{authblk}

\usepackage{amsfonts,amsmath,amssymb}
\usepackage{enumerate}
\usepackage{times}
\usepackage{bm}
\usepackage[percent]{overpic}
\usepackage{float}
\usepackage{graphicx}
\usepackage{lipsum}
\usepackage{bm}
\usepackage[utf8]{inputenc}
\usepackage{amssymb}
\usepackage[justification=centering]{caption}
\usepackage{mathrsfs}
\usepackage{graphicx}
\usepackage{mathtools}
\usepackage{booktabs}
\usepackage{amstext} 
\usepackage{epstopdf}
\usepackage{array} 
\usepackage{booktabs}
\usepackage{caption}
\usepackage{epstopdf}
\usepackage{xcolor}
\usepackage{comment}

\makeatletter

\newcommand{\ba}{ {\boldsymbol a} }
\newcommand{\bA}{ {\boldsymbol A} }

\newcommand{\bB}{ {\boldsymbol B} }

\newcommand{\bd}{ {\boldsymbol d} }

\newcommand{\bH}{ {\boldsymbol H} }

\newcommand{\bN}{ {\boldsymbol N} }

\newcommand{\bQ}{ {\boldsymbol Q} }

\newcommand{\bs}{ {\boldsymbol s} }

\newcommand{\bw}{ {\boldsymbol w} }

\newcommand{\bx}{ {\boldsymbol x} }

\newcommand{\by}{ {\boldsymbol y} }

\newcommand{\bz}{ {\boldsymbol z} }

\newcommand{\bone}{ {\bf 1} }
\newcommand{\bzero}{ {\bf 0} }

\newcommand{\bphi}{ {\boldsymbol \phi} }

\newcommand{\bmu}{ {\boldsymbol \mu} }

\newcommand{\btheta}{ {\boldsymbol \theta} }

\newcommand{\Var}{\mathrm{Var}}
\begin{document}

\begin{center}
    Calibrating satellite maps with field data for improved predictions of forest biomass \\ \vspace{20pt}

    Paul B. May \\
    South Dakota School of Mines \& Technology \\
    Department of Mathematics \\
    501 E St Joseph St, Rapid City, SD, 57701 \\
    paul.may@sdsmt.edu\\ \vspace{20pt}

    Andrew O. Finley \\
    Michigan State University\\
    Department of Forestry\\
    Department of Statistics and Probability\\
    East Lansing, MI \\
    
\end{center}

\noindent\textbf{Abstract:} Spatially explicit quantification of forest biomass is important for forest-health monitoring and carbon accounting. Direct field measurements of biomass are laborious and expensive, typically limiting their spatial and temporal sampling density and therefore the precision and resolution of the resulting inference. Satellites can provide biomass predictions at a far greater density, but these predictions are often biased relative to field measurements and exhibit heterogeneous errors. We developed and implemented a coregionalization model between sparse field measurements and a predictive satellite map to deliver improved predictions of biomass density at a 1 $
\text{km}^2$ resolution throughout the Pacific states of California, Oregon and Washington. The model accounts for zero-inflation in the field measurements and the heterogeneous errors in the satellite predictions. A stochastic partial differential equation approach to spatial modeling is applied to handle the magnitude of the satellite data. The spatial detail rendered by the model is much finer than would be possible with the field measurements alone, and the model provides substantial noise-filtering and bias-correction to the satellite map.

\newpage

\section{Introduction}

Informed forest management and conservation decisions are critical for promoting healthy ecosystems and offsetting carbon emissions through sequestration \citep{houghton2009importance, mo2023integrated}. Aboveground biomass density (AGBD) is a particularly important forest variable, conveying the amount of tree matter and, indirectly, the amount of carbon stored in forests. However, measurement is a prerequisite to management, and inventory of forest resources at actionable scales remains a primary challenge. Forest inventory has traditionally occurred through dispersed samples of \textit{in situ} fields plots. Important forest attributes are measured within the field plots and then the collective sample is used to estimate area averages and totals. Field plot measurements are both expensive and labour intensive, limiting the spatial and temporal density of their collection and therefore the resolution and precision of the resulting estimates \citep{wiener2021united}.\par
The Forest Inventory and Analysis (FIA) program of the US Forest Service manages the national forest inventory for the US \citep{bechtold2005enhanced}. The program manages a network of over 300,000 permanent field plots across the contiguous US. The field plots are revisited on a 5 year cycle for the eastern US and a 10 year cycle for the west. While FIA executes one of the largest forest inventory programs in the world, the scope of the program is matched by the area it intends to manage. There is approximately one field plot per 24 $\text{km}^2$, a density that is drastically lowered if measurements are filtered to be contemporary. If only the field measurements are used, this density often precludes precise estimates at the scale of typical land management and ownership \citep{prisley2021needs}.\par
Remote sensing provides an opportunity to supplement field plots by collecting data correlated with the field-measured attributes at a far greater density than the plots themselves. The Global Ecosystem Dynamics Investigation (GEDI) is a light detection and ranging (lidar) instrument onboard the International Space Station \citep{dubayah2020global}. Specifically designed to measure forest structure, GEDI collects dense but discrete sample measurements along ground tracks as the space station orbits. A primary product of the GEDI mission is the Level 4B (L4B) product, which provides predictions of AGBD at $1 \times 1$ km (hereafter we will simply write 1 km) resolution \citep{l4b}. The L4B predictions are constructed using the ground tracks intersecting the 1 km cell along with regression models to predict AGBD from the observed lidar measurements \citep{patterson2019statistical}.\par 
There is not necessarily a one-to-one relationship between the AGBD predicted by L4B and the AGBD measurements of FIA due to differing field protocols, geographically limited training data for the L4B regression models and potentially other sources of relative bias \citep[Section 3.3]{dubayah2022gedi}. However, the L4B predictions are strongly correlated with FIA-measured biomass, and while not spatially complete, (some 1 km cells have no intersecting ground tracks) are available at a far greater density than the FIA plots. A model fusing both data sources could yield predictions consistent with FIA measurements but with higher precision and at finer resolutions than possible with the plot data alone.\par
The characteristics of the data present a number of challenges to such a model. First, the L4B predictions represent a 1 $\text{km}^2$ area, approximately 100 times the size of the FIA plots (each FIA plot represents approximately 1 hectare). Second, the L4B predictions exhibit substantial heteroskedasticity due to variability in the number of GEDI ground tracks intersecting each 1 km cell. Cells with fewer ground tracks can be expected to exhibit larger sampling error, variation not correlated with the underlying AGBD. A model incorporating L4B but ignoring this heteroskedasticity is liable to be locally under- or over-confident, depending on sampling variance of the L4B prediction. Third, the FIA plot data are ``zero-inflated'': many FIA plots are within non-forested areas and therefore have AGBD measurements of zero \citep{finley2011hierarchical}. The spatial distribution of forest/non-forest is not known \textit{a priori}, therefore these zero-measurements should not be omitted, but rather used within the model to predict non-forested areas and subsequently zero AGBD. The final challenge is only indirectly addressed in this work: the exact FIA plot coordinates are confidential to prevent tampering and maintain land-owner privacy. The public coordinates, used in this study, are perturbed within a kilometer of their true location. \par
We extend the work of \cite{berrocal2010spatio, berrocal2012space}, where a downscaler model was used to link point-referenced direct observations of pollutants to gridded computer model outputs. In the downscaler model, the point observations are associated with the containing grid cell. Of particular importance to our application is the contribution in \cite{berrocal2012space}, where the gridded auxiliary data are not directly linked to the target variable, but linked through a latent mean. This has the potential to be crucial, as the L4B values can be noisy, depending on the number of ground tracks used to produce the prediction. The number of ground tracks per cell is provided by the L4B data set, giving an additional variable that could be used to determine the expected dispersion around the latent mean.\par
We modeled the data with a joint zero-inflated gamma (JZIG) model. The L4B predictions and FIA forested plot observations of AGBD are both modeled with generalized linear models (GLMs) with gamma likelihoods to account for the extremely skewed and strictly positive data. Spatial associations are governed by latent Gaussian processes, where the FIA and L4B models share a Gaussian process to allow dependence. The gamma distribution for L4B has a dispersion parameter that is allowed to vary with the number of ground tracks within a 1 km cell. Finally, a separate Bernoulli GLM is used to account for the zero-inflation in the FIA plot data, dictating whether a location is forest (positive AGBD) or non-forest (zero AGBD). The Bernoulli GLM also shares a Gaussian process with the L4B gamma GLM so that the L4B product informs both the probability of forest presence and, if present, the magnitude of forest AGBD. We used a Bayesian approach to model inference, using Laplace approximations for Gibbs updates of the latent effects. To accommodate the magnitude of the spatial data, we used the stochastic partial differential equation (SPDE) approach \citep{lindgren2011explicit}.\par
We demonstrate this model for the Pacific states of the US (Washington, Oregon, California), producing predictions of AGBD for the year 2019. The JZIG model renders spatial variation of AGBD at far finer scales than would be possible with the plot data alone due to the induced connection with the L4B product. We found the varying dispersion parameter, accounting for the varying number of ground tracks comprising the L4B predictions, played a large role in the posterior precision of the latent Gaussian process. We produced a spatially complete 1 km map of AGBD across the Pacific states paired with rigorous uncertainties. When compared to the original L4B map, this new map exhibits substantial differences, providing predictions more coherent with FIA plot measurements. We explored the predictive capacity of the proposed model using a 10-fold cross-validation study, which showed the JZIG model was able to predict holdout FIA plot measurements well. Finally, we conducted a Monte Carlo study to assess the effect of location uncertainty in the FIA public coordinates for model fitting. Results from the Monte Carlo study suggested that, relative to other sources of uncertainty, there was negligible information loss due to FIA's location perturbation.

\section{Data}

We conducted our analysis over the Pacific states of Washington, Oregon and California, comprising a total area of $8.6 \times 10^5 \text{ km}^2$. We used FIA plot measurements as ``ground-truth" observations of AGBD. A single FIA plot consists of four clustered subplots, representing a total area of approximately 0.1 $\text{km}^2$. We restricted to plots measured in the year 2019 in order to produce baseline predictions of AGBD for a fixed period of time. This resulted in 4,420 plots dispersed across our study area, or approximately one plot per 195 $\text{km}^2$. Almost half of these plot measurements were conducted over a non-forested area and therefore have an AGBD value of zero.\par
To supplement the plot measurements we used the GEDI L4B product, which provides predictions of AGBD at a 1 km resolution. The L4B product uses regression equations trained between simulated GEDI measurements and a crowd-sourced set of field plots measurements, none of which come from the FIA plot network \citep{duncanson2022aboveground}. Predictions at 1 km were produced from the average predictions of intersecting orbital ground-tracks. The predictions were generated using GEDI measurements from April 2019 to March 2023, so there is some temporal mismatch between L4B and the FIA plots. The coverage of the ground-tracks never fills a 1 km cell completely, and we expected more variability in predictions generated from fewer tracks. Some cells have no prediction (around 3.5\% of the study area) either due to orbital patterns or the filtering of low-quality GEDI observations. While the L4B values are regression predictions, the training data are disjoint from the FIA plots, therefore we treated the values simply as observations of a stochastic spatial process, assuming that the dispersion of these observations around an unknown mean will be related to the number of ground-tracks within a cell. In total, there are 833,538 L4B observations within our study area. Around 4.3\% of the L4B observations of AGBD are exactly zero. Rather than introduce an additional modeling component to account for a modest number of zeros, we set these observations to the minimum of the positive observations, $0.003$ Mg/ha.


\begin{figure}
    \centering
    \includegraphics[width = \textwidth, keepaspectratio]{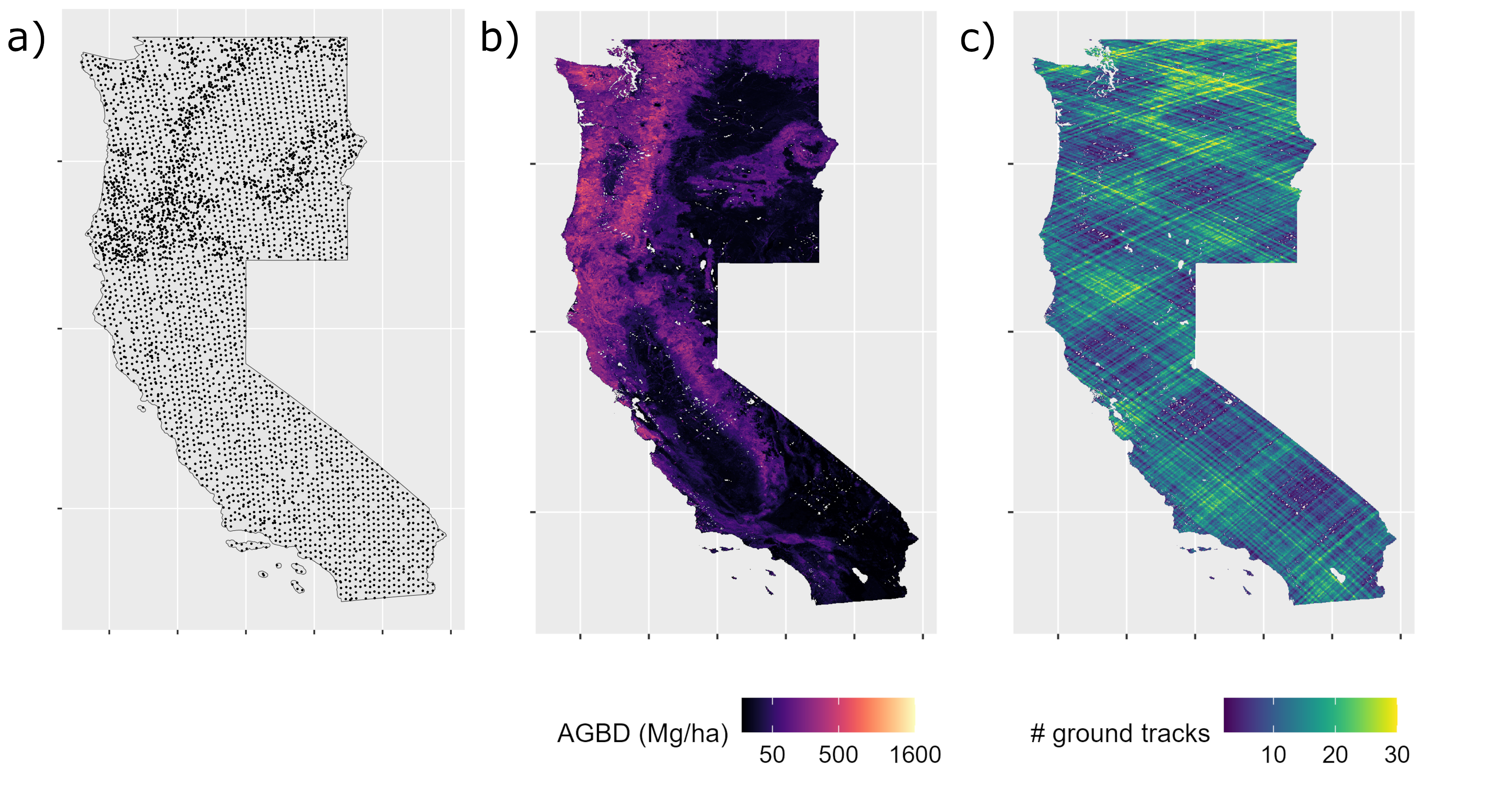}
    \caption{Spatial maps of the data. Panel (a) gives the locations of spatially perturbed FIA plots. Panel (b) gives the AGBD values of the L4B product, while (c) gives the number of ground tracks used to produce those values.}
    \label{fig:data}
\end{figure}

\section{Methods}\label{sec:methods}
Section \ref{sec:prelim} gives an introduction to the Gaussian processes used in this work and their SPDE representations. Section \ref{sec:model} details the JZIG model, Section \ref{sec:inference} gives an overview of the Bayesian inference and Section \ref{sec:spde} gives more specific details on our implementation of the SPDE approach.

\subsection{Preliminaries}\label{sec:prelim}
Over the past two decades, Gaussian processes have become central to spatial statistics, providing a prior for random effects that can account for spatial associations in data while being analytically tractable due to the closed form conditional distributions of the multivariate normal family \citep{gelfand2016spatial}. In this work, we used Gaussian processes with the Mat\'ern covariance function of smoothness $\nu = 1$. Let $\eta(\bs)$ be a Gaussian process with domain $\bs\in \mathcal{D}$ and Mat\'ern covariance. Then
\begin{equation}
    \mathrm{Cov}\left[\eta(\bs), \eta(\bs')\right] = \sigma^2\sqrt{8}\frac{\|\bs-\bs'\|}{\rho}K_1\left(\sqrt{8}\frac{\|\bs-\bs'\|}{\rho}\right) \label{eq:matern}
\end{equation}
for $s,s'\in \mathcal{D}$, where $K_1(\cdot)$ is an order $1$ modified Bessel function of the second kind. The Mat\'ern covariance function is governed by a standard deviation parameter $\sigma$, giving the prior standard deviation for $\eta(\bs)$, and a range parameter, $\rho$, giving (inversely) the rate at which the correlation between $\eta(\bs)$ and $\eta(s')$ decays as the distance between the locations, $\|s - s'\|$, increases. We used the parameterization of \cite{lindgren2011explicit} in (\ref{eq:matern}), where $\rho$ is interpretable as the distance for which the correlation is approximately 0.13.\par
For large sample sizes, using Gaussian processes natively is infeasible, as evaluating the prior log-density requires the solution of $n\times n$ dense linear systems and log-determinants, where $n$ is the number of observations. We instead used the SPDE approach \citep{lindgren2011explicit}. The SPDE approach arises from the acknowledgement that a Gaussian process with Mat\'ern covariance is the stationary solution to an SPDE. This solution admits a finite element representation, where the finite elements are given by a \lq mesh\rq\ partitioning a greater domain $\mathcal{D}^*$ into disjoint triangles. The greater domain $\mathcal{D}^*$ serves to contain and buffer target domain, $\mathcal{D}\subset \mathcal{D}^*$, to avoid boundary effects from the imposed SPDE boundary conditions. A Gauss-Markov random field (GMRF), $\bw = [w_1~\cdots,w_k]^T$, is defined on the $k$ triangle vertices with zero mean and sparse precision matrix $\bQ(\sigma, \rho)$. To extend the GMRF to the continuous domain, $k$ piece-wise linear basis functions are defined, where each basis function is compactly supported within an assigned triangle. Then $\eta(\bs)$ is defined by
\begin{equation}
    \eta(\bs) = \ba(\bs)^T\bw;~~~~~\ba(\bs) = \begin{bmatrix} a_1(\bs) \\ \vdots \\ a_k(\bs) \end{bmatrix},
\end{equation}
where $a_j(\bs);~j\in{1,\ldots,k}$ are the basis functions, with at most three non-zero evaluations for any given $s\in\mathcal{D}$. The quality of this representation depends on the edge-length of the mesh triangles relative to the Mat\'ern range, $\rho$. For a good representation, the edge-lengths should be substantially shorter than $\rho$ so that adjacent $w_i,w_j$ are highly correlated. Otherwise, the mesh triangles will be visible in predictive inference and $\eta(\bs)$ will provide a poor model for a continuous spatial field.

\subsection{Model}\label{sec:model}
Let $x(B) > 0$ be the L4B biomass prediction at arbitrary 1 km cell $B$ and $N(B)$ be the number of intersecting ground tracks. We assumed the GLM
\begin{align}
    x(B) &\sim \text{Gamma}\left(\mu_x(B),~ \phi_x + \frac{\phi_g}{N(B)}\right) \label{eq:xrealize}\\
    \log(\mu_x(B)) &= \alpha_x + \eta_x(B) \label{eq:mux}.
\end{align}
Within the latent mean process, $\mu_x(B)$, parameter $\alpha_x$ is an intercept term and latent effect $\eta_x(B)$ is a Gaussian process with zero mean and Mat\'ern covariance, where the distance in (\ref{eq:matern}) is evaluated between the cell centers. Because the L4B predictions occur on a regular grid, it would be natural to assign an autoregressive structure to $\eta_x(B)$, as done in \cite{berrocal2012space}, rather than the spatially-continuous covariance function. However, the rank of the resulting precision matrix for all L4B predictions would be almost $10^6$, an intimidating prospect even with sparse matrix routines. If the scale of spatial variation is substantially larger than 1 km, then the SPDE approach will allow a lower-rank representation. Parameters $\phi_x,~\phi_g > 0$ are dispersion parameters, where for fixed mean process,
\begin{equation}\label{eq:varyingdispersion}
    \Var[\,{x(B)}\,|\,\mu_x(B),\, \phi_x,\, \phi_g\,] = \left(\phi_x + \frac{\phi_g}{N(B)}\right)\mu_x(B)^2,
\end{equation}
giving a portion of the variance, $\phi_g$, that decreases with the number of intersecting ground tracks, $N(B)$. If $\phi_g$ is inferred to be of substantial magnitude, the dispersion of $x(B)$ around $\mu_x(B)$ will decrease with the number of ground tracks, allowing more precise recovery of the latent effect $\eta_x(B)$ around cells with many orbits.\par
Let $y(\bs)\geq 0$ be an FIA plot measurement at location $s$. The FIA measurements of AGBD are zero-inflated, with a large fraction of zero AGBD measurements due to non-forested plots. Thus, we assume a Gamma GLM similar to (\ref{eq:xrealize}, \ref{eq:mux}), but with a Bernoulli GLM (logistic regression) to separate the forest and non-forest locations. For location $s$ within cell $B$, 

\begin{align}
    z(\bs) &\sim \mathrm{Bernoulli}\left(\mu_z(\bs)\right) \\
    \mathrm{logit}(\mu_z(\bs)) &= \alpha_z + \eta_z(\bs) + \beta_z\eta_x(B) \label{eq:muz} \\
    y(\bs)\, |\, \left( z(\bs) = 1\right) &\sim \text{Gamma}(\mu_y(\bs), \phi_y)\\
    \log(\mu_y(\bs)) &= \alpha_y + \eta_y(\bs) + \beta_y\eta_x(B) \label{eq:muy} \\ 
    y(\bs)\, |\, \left( z(\bs) = 0 \right) &= 0 .
\end{align}
Parameters $\alpha_z,~\alpha_y$ are intercept terms, $\phi_y$ is a constant dispersion parameter, and effects $\eta_z(\bs)$ and $\eta_y(\bs)$ are point-referenced Gaussian processes with zero mean and Mat\'ern covariance. The L4B map informs both the probability of forest presence and, given forest presence, the AGBD magnitude through the terms $\beta_z\eta_x(B)$ and $\beta_y\eta_x(B)$, where $\beta_z$ and $\beta_y$ are regression coefficients. Thus, the plot measurements are only connected to L4B through latent, spatially smooth terms, such as in \cite{berrocal2012space}. We hypothesized the spatial scale of variation in $\eta_x(B)$ to be broad enough (larger than 1 km) to ameliorate any inconsistencies from the change of support (1 $\text{km}^2$ L4B cells linked to 0.1 $\text{km}^2$ plot measurements) or from the plot coordinates only being known up to a kilometer.\par
If the variances of $\beta_z\eta_x(B)$ and $\beta_y\eta_x(B)$ are substantial relative to the variances of $\eta_z(\bs)$ and $\eta_y(\bs)$, then the inclusion of L4B will be highly informative on AGBD, $y(\bs)$. Further, since the L4B predictions are given on a 1 km grid while there is on average one FIA plot per 195 $\text{km}^2$, we hypothesized that $\eta_x(B)$ would account for finer scale spatial patterns, leaving broad spatial patterns to $\eta_y(\bs)$

\subsection{SPDE Implementation}\label{sec:spde}
There are over $4\times 10^3$ and $8\times 10^5$ observations from the plots and L4B, respectively, so modeling directly with Gaussian processes would be flatly impossible. We therefore modeled all Gaussian processes with the SPDE approach described in Section \ref{sec:prelim}.\par
Expecting different scales of variation between the processes, we assigned a mesh to $\eta_x(\bs)$ and a separate mesh to $\eta_y(\bs)$ and $\eta_z(\bs)$: the L4B cells are observed at a far greater density and therefore have the potential to capture finer spatial patterns. Through a few iterations of model training, we decreased the maximum-allowed edge-length for the meshes until they were less than $1/5$ of the posterior expected Mat\'ern range for the processes(es) they represent. The resulting mesh for $\eta_x(\bs)$ had a maximum edge-length of 2 km, giving $k_x =\, $162,328 vertices, while the mesh for $\eta_y(\bs), \eta_z(\bs)$ had a maximum edge-length of 20 km, giving $k_y=k_x=\,$6,862 vertices.
\begin{figure}
    \centering
    \includegraphics[width = \textwidth, keepaspectratio]{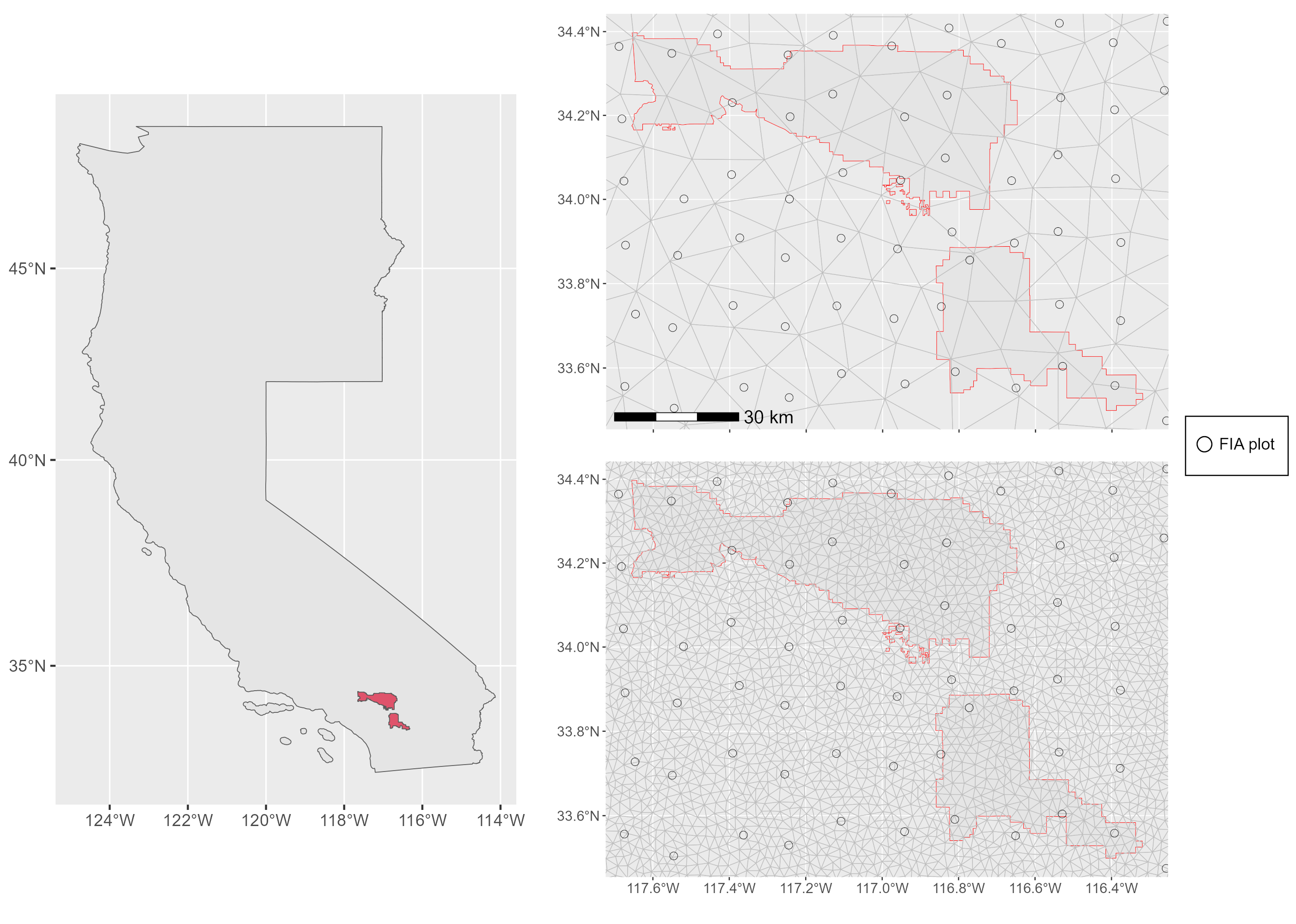}
    \caption{The Gaussian processes are represented as projections of a Gauss-Markov random field on a mesh. The left panel gives the Pacific states with San Bernardino National Forest highlighted in the south. The top-right panel gives a zoomed view of the coarser mesh for $\eta_y(\bs)$ and $\eta_z(\bs)$ over this national forest, while the bottom-right panel gives the much finer mesh for $\eta_x(\bs)$. FIA plots are plotted with circles of radius 1 km, depicting the uncertainty of the plots' true location.}
    \label{fig:meshes}
\end{figure}
The basis functions depend only on the mesh, giving representations
\begin{equation}
    \eta_x(B) = \ba_x(B)^T\bw_x;~~~\eta_y(\bs) = \ba(\bs)^T\bw_y;~~~\eta_z(\bs) = \ba(\bs)^T\bw_z,
\end{equation}
where $\ba_x(B)$ are the basis functions evaluated at the geographic center of the 1 km cell.

\subsection{Bayesian Inference}\label{sec:inference}

We assigned flat priors to the regression-like parameters $\alpha_x,~\alpha_y,~\alpha_z$ and $\beta_y,~\beta_z$. Flat priors on the log-scale were assigned to the dispersion parameters $\phi_x,~\phi_g,~\phi_y$. The Mat\'ern covariance parameters are not consistently estimable without informative priors. Therefore, we used penalized complexity priors for a joint prior on each of the standard deviation/range pairs \citep{fuglstad2019constructing}. This prior constrains the values by placing upward pressure on the range and downward pressure on the standard deviation. Further, the prior parameters can be specified using interpretable probability statements. We gave the prior assumptions:

\[\begin{array}{ll}
    \mathrm{Prob}(\sigma_x > 2) = 0.01 & \mathrm{Prob}(\rho_x < 9 \text{ km}) = 0.01 \\
    \mathrm{Prob}(\sigma_y > 0.5) = 0.01 & \mathrm{Prob}(\rho_y < 60 \text{ km}) = 0.01 \\ 
    \mathrm{Prob}(\sigma_z > 1) = 0.50 & \mathrm{Prob}(\rho_z < 60 \text{ km}) = 0.01
\end{array}\]

Processes $\eta_y(\bs)$ and $\eta_z(\bs)$ describe residual variation in the plot measurements not shared by L4B. Previous studies have found the differences between the two to primarily vary across large distances \citep{dubayah2022gedi}. We therefore assigned little prior probability that the range of these processes would be less than $60 \text{ km}$. The L4B is observed at a much greater density and has the potential to capture finer spatial variation, so the range of $\eta_x(B)$ was given more prior allowance to take lower values.\par
We used a Gibbs sampler to generate approximate samples from the posterior distribution. The samples are approximate because Laplace approximations were employed to generate conditional samples of the latent effects within the mean processes of (\ref{eq:mux}), (\ref{eq:muz}) and (\ref{eq:muy}), as no closed-form Gibbs updates are available for these effects. For example, we use a Laplace approximation for the block-Gibbs update of $\btheta_x = [\alpha_x~\bw_x^T]^T$: a Newton-Raphson procedure is used to find the conditional posterior mode, $\hat{\btheta}_x$. Then a sample is drawn as $\btheta_x^{(k+1)} \sim \mathrm{MVN}\left(\hat{\btheta}_x, \bH(\hat{\btheta}_x)^{-1}\right)$, where $\bH(\btheta_x)$ is the Hessian of the log-conditional evaluated at $\btheta_x$. Each Newton-Raphson iteration and the subsequent simulation from the multivariate normal distribution evaluates a Cholesky decomposition of the Hessian. The Hessian is large, of dimension 1 + 162,328 in the case of $\btheta_x$, but extremely sparse due to the sparse prior precision of $\bw_x$. Therefore these Cholesky decompositions are feasible with sparse matrix methods. Using the CHOLMOD library \citep{cholmod}, a decomposition of $\bH(\btheta_x)$ required $\sim7$ seconds on a 1.8 GHz commercial laptop. Each full iteration of the Gibbs sampler required 25 -- 50 seconds on this same machine, depending on the number of iterations required for the Newton-Raphson procedure to converge: in our experience, as the MCMC chain approached the stationary distribution, the number of Newton-Raphson iterations required for convergence decreased and then stabilized.  The Gibbs sampler and Laplace approximations are described in detail in Appendix \ref{app:GibbsNLaplace}.\par
Using the posterior samples of the model parameters and effects, it is straightforward to generate predictive samples of $y(\bs)$ at unobserved locations:
\begin{align}
    \mathrm{logit}\left(\mu_z^{(k)}(\bs)\right) &= \alpha_z^{(k)} + \ba(\bs)^T\bw_z^{(k)} + \beta_z^{(k)}\ba_x(B)^T\bw_x^{(k)}, \\
    z^{(k)}(\bs) &\sim \mathrm{Bernoulli}\left(\mu_z^{(k)}(\bs)\right), \\
    \log\left(\mu_y^{(k)}(\bs)\right) &= \alpha_y^{(k)} + \ba(\bs)^T\bw_y^{(k)} + \beta_y^{(k)}\ba_x(B)^T\bw_x^{(k)}, \\
    y^{(k)}(\bs)\, |\, \left(z^{(k)}(\bs) = 1\right) &\sim \mathrm{Gamma}\left(\mu_y^{(k)}(\bs),~\phi_y^{(k)}\right), \\
    y^{(k)}(\bs) &=  z^{(k)}(\bs)\cdot\left[y^{(k)}(\bs)\, |\, \left(z^{(k)}(\bs) = 1\right)\right]
\end{align}
where $k$ is the given MCMC sample, $B$ is the cell containing $s$. Note that $\eta_x(B),~\eta_z(\bs)$ and $\eta_y(\bs)$ are replaced with their SPDE representations, as this is how predictions will be conducted in practice. Inference on AGBD is typically desired for areas larger than a plot, such as a contiguous forest stand, a fireshed or an ownership tract. Let $A$ be an area for which a prediction is desired. Variable $y(\bs)$ does not represent the AGBD at an infinitesimal point, but rather the AGBD for a hypothetical 0.1 $\text{km}^2$ plot centered at $s$. Assume that area $A$ can be partitioned into disjoint plots at $s_1, s_2, \ldots s_p$, where $p$ is the number of disjoint plots. Then a predictive sample of the area AGBD is given by
\begin{equation}
    y^{(k)}(A) = \frac{1}{p}\sum_{i=1}^p y^{(k)}(s_i).
\end{equation}
In this work, we replicated the resolution of L4B map, giving 1 $\text{km}^2$ predictions across the Pacific states. 

\section{Data Analysis}\label{sec:results}

Given the data, we generated the approximate posterior distribution for the JZIG model unknowns using the methods in Section \ref{sec:inference} and Appendix \ref{app:GibbsNLaplace}. Process $\eta_x(\bs)$, the latent variable shared by both L4B and the plot measurements, was found to account for a large fraction of the total variance in the plot measurements: Considering the mean process for the positive biomass plots in (\ref{eq:muy}), the posterior expectation of $\beta_y\sigma_x$ was $1.72$, whereas the posterior expectation of $\sigma_y$ was $0.23$. Considering the mean process for the Bernoulli variable in (\ref{eq:muz}), distinguishing forest and non-forest, the posterior expectation of $\beta_z\sigma_x$ was $3.97$, whereas the posterior expectation of $\sigma_z$ was $1.47$. Posterior expectations and 95\% credible intervals for all model parameters are given in Table \ref{tab:params}. \par

\begin{table}[]
    \caption{Posterior expected values and 95\% credible intervals (equal-tail) for the model parameters.}
    \centering
    \begin{subtable}[t]{\textwidth}
        \caption{Parameters for $x(B)$ model}
        \begin{tabular}{c c c c c c}\toprule
          Parameter & $\alpha_x$ & $\sigma_x$ & $\rho_x$ (km) & $\phi_x$ & $\phi_g$  \\ 
          Expectation & 2.75 & 1.87 & 16.7 & 0.050 & 3.81 \\
          95\% CI & (2.67, 2.82) & (1.85, 1.89) & (16.4, 16.9) & (0.048, 0.053) & (3.78, 3.83) \\ \bottomrule 
        \end{tabular}
    \end{subtable}
    \begin{subtable}[t]{\textwidth}
        \caption{Parameters for $y(\bs)$ model}
        \begin{tabular}{c c c c c c}\toprule
          Parameter & $\alpha_y$ &  $\beta_{y}$ & $\sigma_y$ & $\rho_y$ (km) & $\phi_y$    \\ 
          Expectation & 2.17 & 0.92 & 0.23 & 194.9 & 0.64 \\
          95\% CI & (1.99, 2.31) & (0.86, 0.97) & (0.20, 0.26) & (163.3, 220.4) & (0.61, 0.67)  \\ \bottomrule 
        \end{tabular}
    \end{subtable}
    \begin{subtable}[t]{\textwidth}
        \caption{Parameters for $z(\bs)$ model}
        \begin{tabular}{c c c c c }\toprule
          Parameter & $\alpha_z$  & $\beta_z$ & $\sigma_z$ & $\rho_z$ (km) \\ 
          Expectation & -2.45 & 2.12 & 1.47 & 167.6 \\
          95\% CI & (-3.11, -1.92) & (1.94, 2.28) & (1.26, 1.71) & (141.7, 192.8) \\ \bottomrule 
    \end{tabular}
    \end{subtable}
    \label{tab:params}
\end{table}

\begin{figure}
    \centering
    \includegraphics[width = \textwidth]{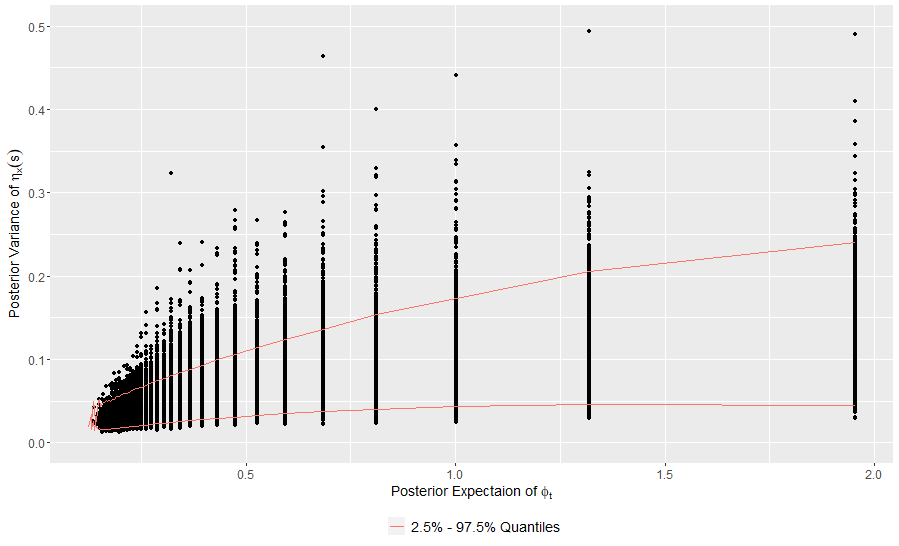}
    \caption{Effect of the varying dispersion parameter, $\phi_t = \phi_x + \phi_g/N(B)$ on the recovery of latent Gaussian process $\eta_x(\bs)$ for the Pacific states. As the number of observed orbits within a cell increase, the dispersion decreases, allowing in general more precise recovery of the latent effect. Points are a random sample of $2\times10^5$ L4B-observed cells, but the quantiles were computed with all $\sim8\times10^5$ L4B observations.}
    \label{fig:varyingdispersion}
\end{figure}

The total dispersion parameter for the L4B observations was found to vary drastically across the 1 km cells, having a substantial impact on the posterior precision of $\eta_x(\bs)$ (Fig. \ref{fig:varyingdispersion}). The posterior expectations of $\phi_g$ and $\phi_x$ where approximately $3.81$ and $0.05$ respectively. Recall that the total dispersion was defined as $\phi_t(B) = \phi_x + \phi_g/N(B)$. Therefore, the total dispersion, and correspondingly the recovery of $\eta_x(B)$, was heavily influenced by the number of ground tracks within a 1 km cell, $N(B)$. \par

\begin{figure}
    \centering
        \includegraphics[width = \textwidth, keepaspectratio]{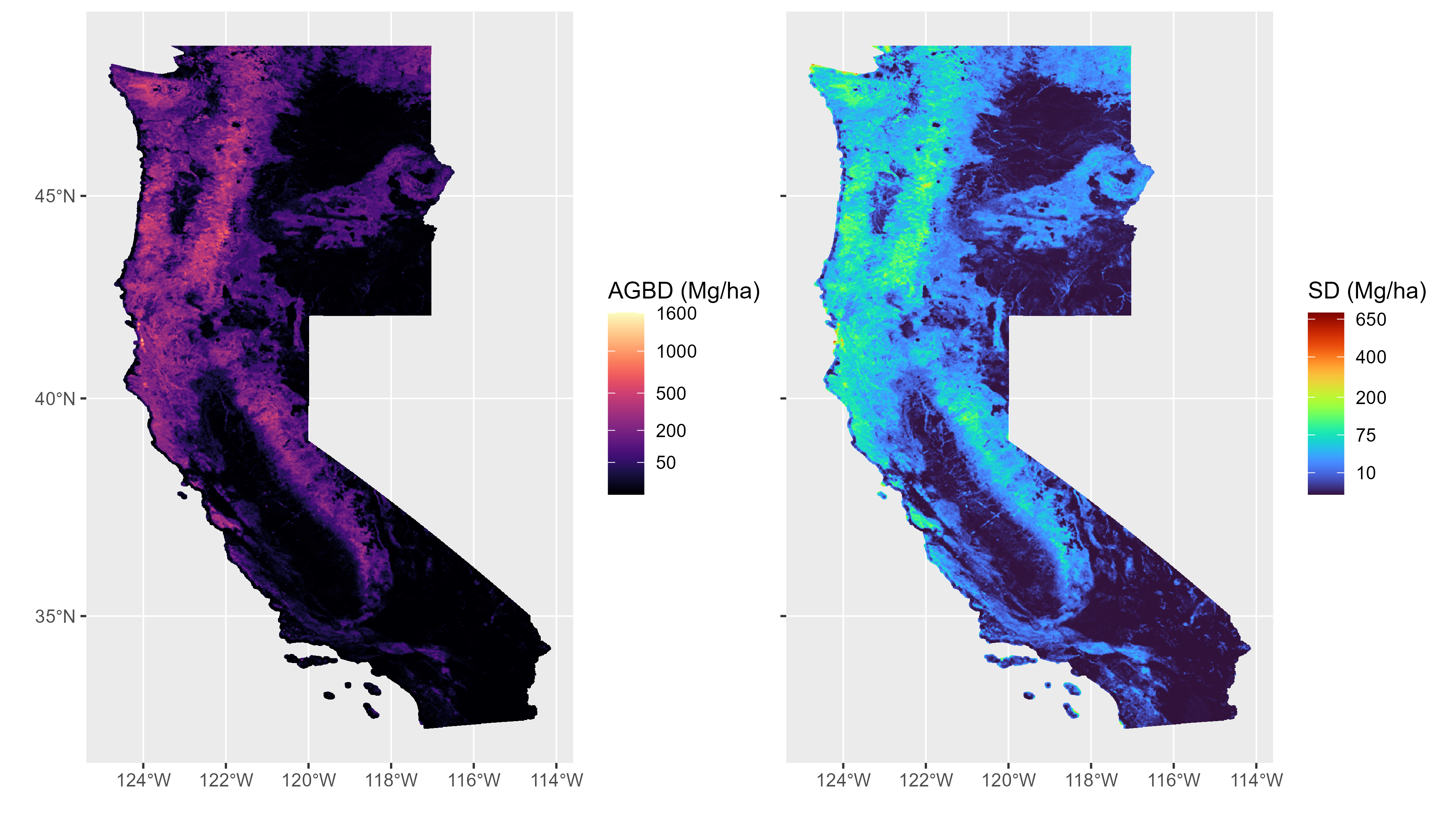}
    \caption{Expected values and standard deviations for forest AGBD at 1 km resolution for the Pacific states.}
    \label{fig:pred_km}
\end{figure}

\begin{figure}
    \centering
    \includegraphics[width = \textwidth]{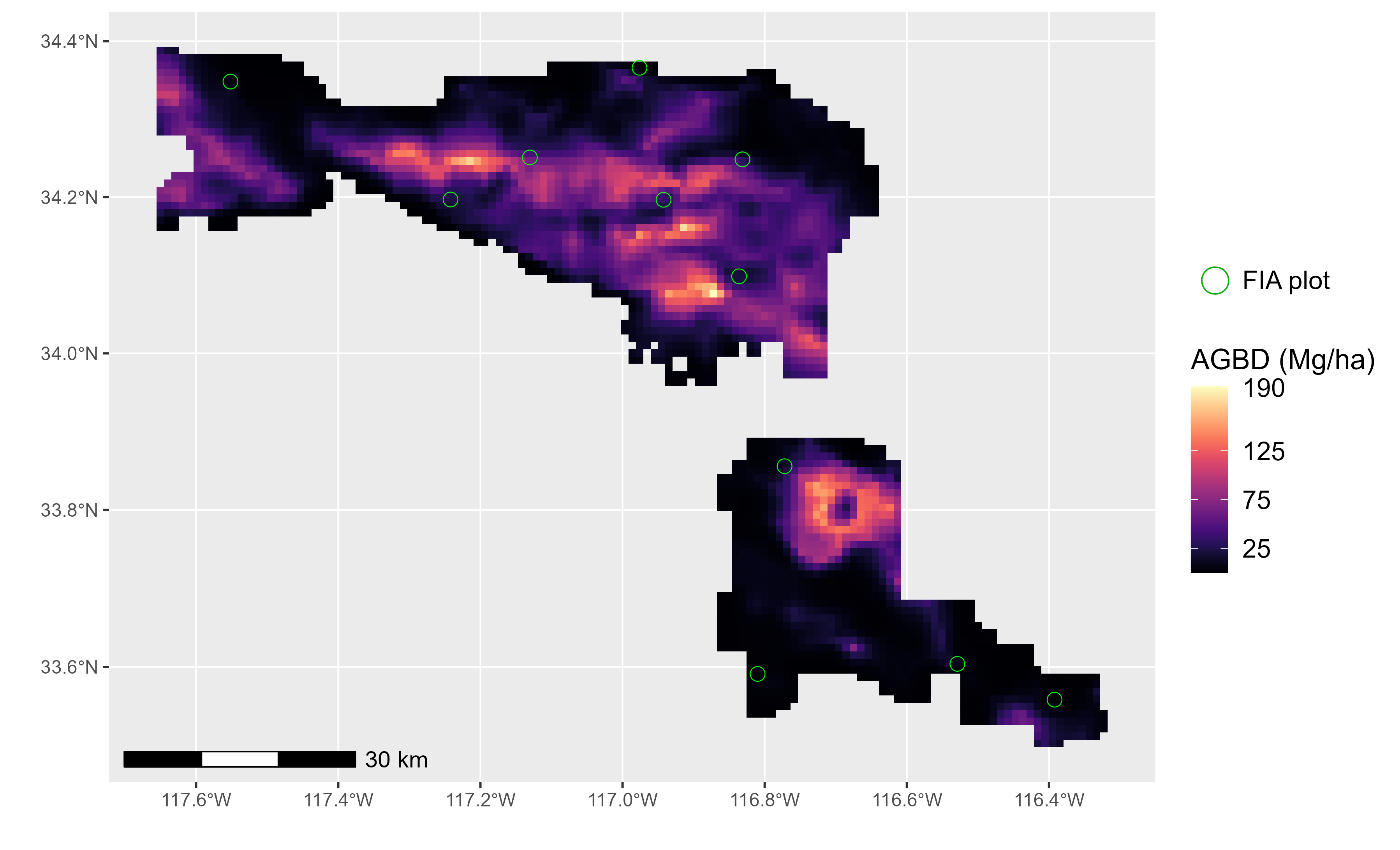}
    \caption{Posterior expected AGBD at 1 km resolution for San Bernardino National Forest. FIA plots are plotted with circles of radius 1 km, depicting the uncertainty of the plots true location. Within this area, the spatial variation of AGBD occurs at scales finer than the density of plots, leaving many of the highest concentrations of biomass unrepresented in the plot sample.}
    \label{fig:pred_km_sb}
\end{figure}

We generated posterior predictions of AGBD, $y(\bs)$, at the 1 km resolution of L4B across the Pacific states (Fig. \ref{fig:pred_km}). Focusing on San Bernardino National Forest, the scale of spatial variation captured by the predictive map is far finer than what could be inferred using the plot data alone (Fig. \ref{fig:pred_km_sb}). Indeed, entire contiguous stands of forest possessing substantial AGBD are missed by the plot sample altogether. \par

\begin{figure}[htb]
    \centering
    \includegraphics[width = \textwidth]{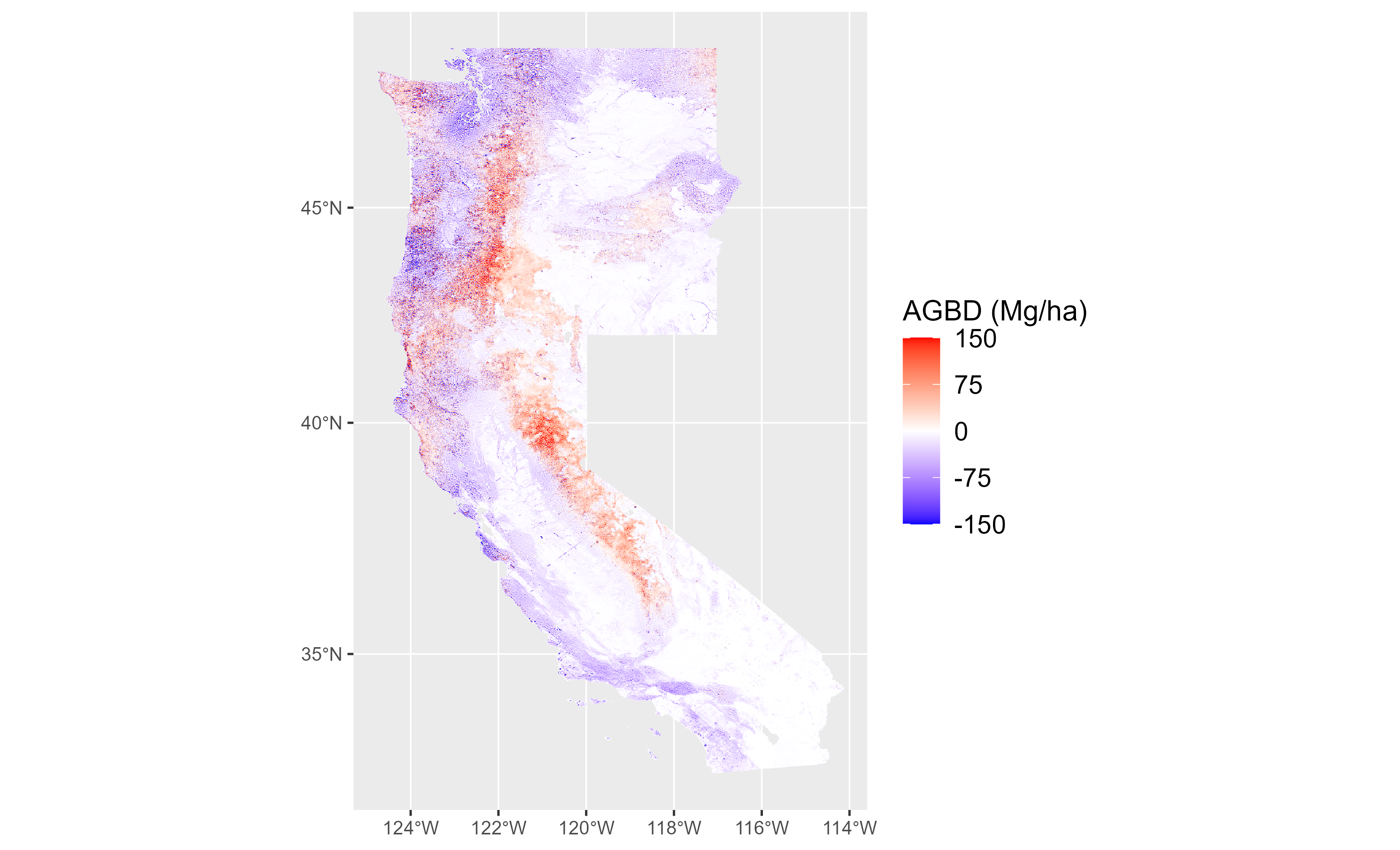}
    \caption{The JZIG expected AGBD at 1 km minus the L4B predictions. Relative to FIA plot measurements, L4B under-predicts AGBD for much of the Sierra Nevada and Cascades, and over-predicts AGBD for many other areas.}
    \label{fig:l4b_diff}
\end{figure}

The JZIG expected AGBD values at 1 km exhibit physically substantial and spatially varying differences from the original L4B predictions (Fig. \ref{fig:l4b_diff}). For example, much of the Sierra Nevada and Cascades mountain ranges have JZIG expectations greater than the L4B predictions by $>100$ Mg/ha. These same relative spatial biases between FIA and L4B were noted in \cite[Fig.~8]{dubayah2022gedi}, and are corrected in the JZIG model, which reconciles the two data sources.\par

We conducted a single iteration of 10-fold cross-validation to assess the performance of the model in predicting the FIA-measured AGBD. After randomizing the plot order, the plots were partitioned into ten folds. Sequentially, each fold of plots was withheld as a test set. The remaining folds were used to train the model and subsequently produce predictive distributions for the withheld plots. Because the AGBD measurements are semi-continuous due to the zero-inflation, we separate the assessment into two parts. In this first part of the validation, the performance of the Bernoulli GLM was examined through a calibration curve, comparing the expected probability of forest presence to the binary forest presence/absence of the test plots. The calibration curve was computed as loess curve between the expected probability and the true binary value. We would expect that the set of predictions with a predicted probability of around 0.3, for instance, should correspond to average binary value of around 0.3. Thus, a well-performing model should yield a calibration curve close to the one-to-one line. Within the second part of the validation, we assessed the coverage of the 95\% credible intervals for $y(\bs)|z(\bs) = 1$ for the forested test plots. Credible intervals were computed as equal-tail intervals.\par
\begin{figure}
    \centering
    \begin{subfigure}[b]{0.49\textwidth}
        \centering
        \includegraphics[width = \textwidth, keepaspectratio]{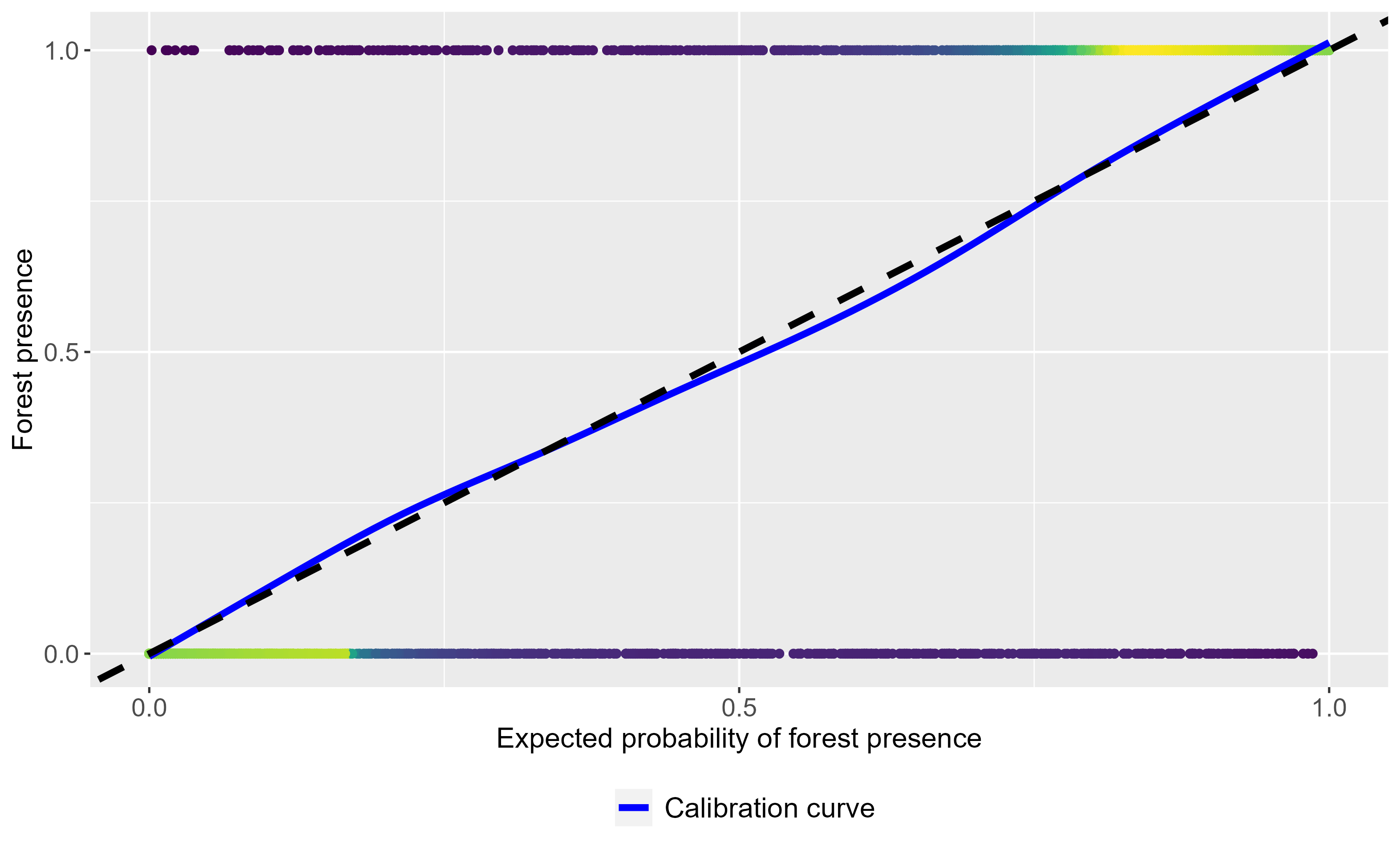}     
    \end{subfigure}
    \hfill
    \begin{subfigure}[b]{0.49\textwidth}
        \centering

        \includegraphics[width = \textwidth, keepaspectratio]{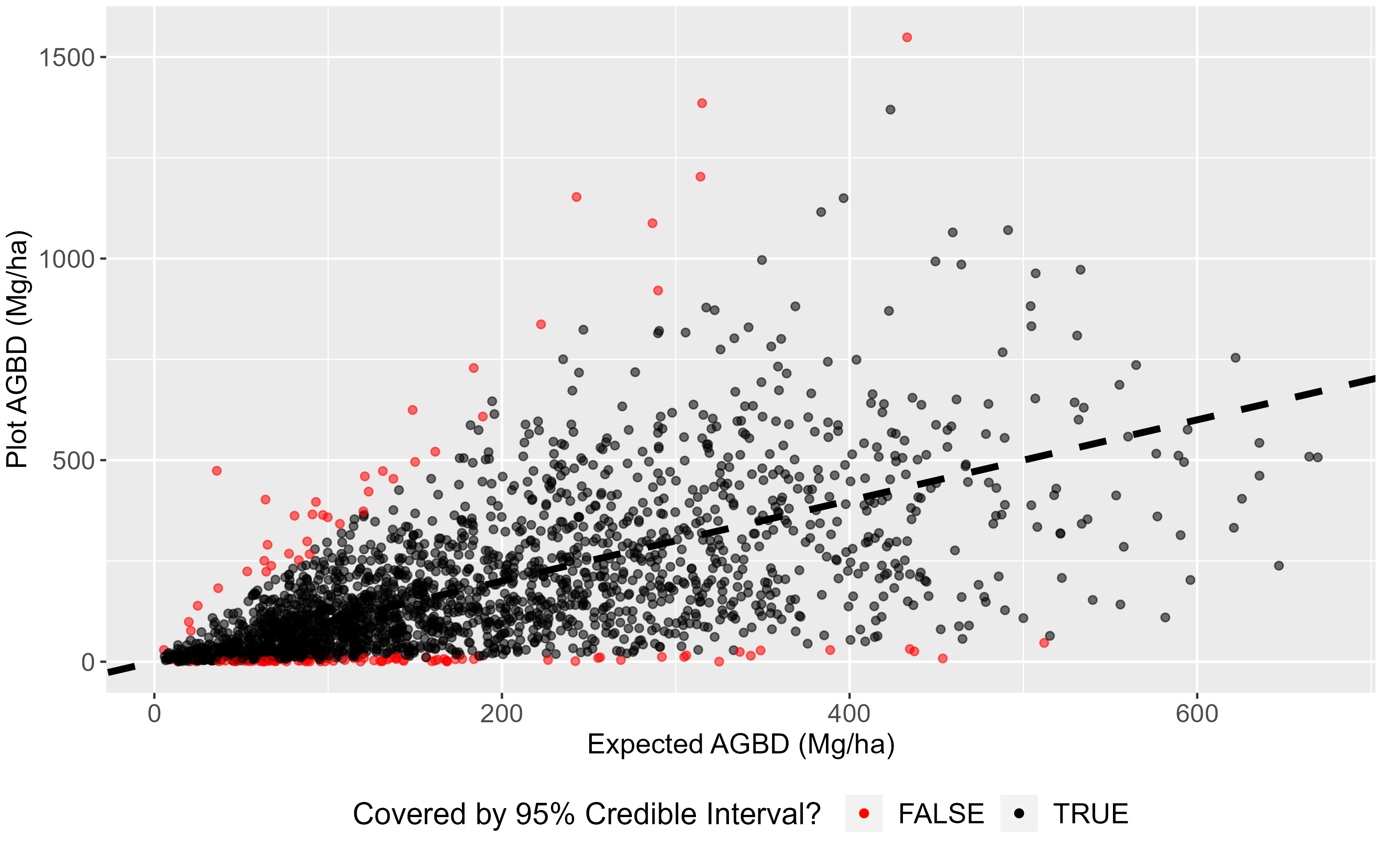}
    \end{subfigure}
    \caption{Results of the cross-validation study. The left panel gives the calibration curve for the Bernoulli GLM, separating forest and non-forest, where points in the top and bottom ribbon are colored by the density of neighboring points. The right panel gives the posterior expected versus actual AGBD values for the forested, positive AGBD plots.}
    \label{fig:cv}
\end{figure}

For both parts of the validation, the model performed well. The calibration curve remains near the one-to-one line, and the coverage rate for the 95\% credible intervals over the forested plots was 94.5\% (Fig. \ref{fig:cv}). Further, a persistent problem arising in AGBD model-based predictions is poor behavior within high biomass plots, where an overall satisfactory coverage rate is actually the result of slight over-coverage for low-to-moderate AGBD plots and drastic under-coverage for high AGBD plots. The JZIG model did not exhibit this, with even coverage across plots with less than 100 Mg/ha and plots with greater than 100 Mg/ha.

Finally, we performed a Monte Carlo study to examine the effect of the uncertain FIA public coordinates (public coordinates are perturbed within a kilometer of their true coordinates). For 100 iterations, new coordinates for every location were simulated according to a uniform distribution on a disk with radius 1 km respectively centered at each public coordinate. Computing full posterior samples 100 times would be onerous. However, only some parameters/effects are expected to be affected \textit{a posteriori} by the reported coordinates, particularly $\beta_z$ and $\eta_z(\bs) = \ba(\bs)^T\bw_z$ within equation (\ref{eq:muz}) and $\beta_y$ and $\eta_y(\bs) = \ba(\bs)^T\bw_y$ within equation (\ref{eq:muy}). Some effect on $\eta_x(\bs) = \ba_x(\bs)^T\bw_x$ could be expected, but at a milder degree, as $\eta_x(\bs)$ will be primarily informed by the dense L4B observations, $x(B)$. Let $\theta_* = \{\beta_z,~\bw_z,~\beta_y,~\bw_y\}$. The remaining parameters effects $\theta\setminus\theta_*$ were fixed at a random posterior sample. Then for each new coordinate simulation, the conditional posterior mode for $\theta_*$ was computed, $\hat{\theta}_*^{(k)};~k=1,\ldots,~100$, which also gives the posterior expected value of the Laplace approximation. We examined the difference between the minimum and maximum mode computed across all 100 simulations. This difference was $1.6 \times 10^{-4}$ for $\beta_z$ and $5.8 \times 10^{-5}$ for $\beta_y$. Across all vector entries of $\bw_z$, the maximum difference was $1.3 \times 10^{-3}$, and across all vector entries of $\bw_y$, the maximum difference was $2.9 \times 10^{-4}$. Relative to the posterior expected values and standard deviations for these parameters (Table \ref{tab:params}), these differences seem indistinguishable from numerical precision errors.   

\section{Discussion}

Forest inventory data collected in the field is often spatially sparse due to the expense of collection. On the other hand, satellite products can provide dense predictions strongly correlated with but locally inconsistent with the field measurements. A coherent probabilistic model can reconcile the two, providing finer-scale predictions on forest attributes that are consistent with the field measurements. Importantly, these predictions are paired with rigorous uncertainties, necessary to make the resulting maps trustworthy to decision makers \citep{mcroberts2011satellite}.\par
Such modeling efforts are not without challenges. Field measurements of biomass are heavily skewed and zero-inflated. The JZIG model successfully separated the forest and non-forested areas within its stated uncertainties (Fig.~\ref{fig:cv}), leaving the continuous but skewed forested AGBD values to be modeled by a Gamma distribution. For sampling missions such as GEDI, the density of observations is far from constant, creating heteroskedasticity in the resulting maps such as L4B that should be accounted for. Indeed, we found the number of intersecting orbits used to generate an L4B 1 km prediction played a substantial role in the recovery of the latent mean (Fig.~\ref{fig:varyingdispersion}). Finally, the resolution of the satellite maps rarely align with that of the field measurements. Linking both sources with latent spatial processes that vary at scales larger than either resolution ameliorates inconsistencies from the differing supports. This technique also appears to provide inference stable to non-exact plot coordinates, which is attractive to inventory programs seeking to maintain confidential coordinates while delivering to stakeholders forest attribute information at scales finer than the density of plots.\par
A number of avenues exist to further this research. First, while the JZIG model using L4B as auxiliary data can provide information at scales finer than the density of the plots, a primary scale of forest management is the forest stand (a contiguous group of trees of similar age, species, etc.) \citep{makela2012stand, yoshimoto2016stand}. Most forest stands are smaller than a square kilometer. It is not sensible that our current model and data could achieve accurate information at the scale of the typical forest stand, as the finest data source, L4B, is observed at a kilometer density. Using the raw GEDI footprint data \citep{l2a, l4a} that generates the 1 km predictions of L4B could refine the scale of information, but would drastically increase the data volume and computational effort.\par
Second, we filtered the FIA plot data to measurements in 2019 to provide a time-static baseline of forest biomass stocks. A critical need will be providing accurate estimates of biomass \emph{change} over land management and ownership units, incorporating historical and contemporary FIA plot data. Such is especially important in California, where a forest carbon offset program has been initiated which financially compensates private entities for maintaining and promoting forests under their ownership. The model implemented in this work could be adapted for spatial-temporal random effects, such as in \cite{berrocal2012space}. A challenge is that while FIA's current plot sampling regime dates back to approximately 2001 \citep{bechtold2005enhanced}, satellite measurements sensitive to forest structure such as GEDI are modern (GEDI began collection in 2019). An ideal model would efficiently incorporate the contemporary satellite data, which occupies a narrow temporal band relative to the field plot collection history. Great care would be required for model specification pertaining to satellite-to-field measurement relationship, as model-induced bias could predict spurious change at the beginning of the satellite collection period.  

\section*{Data Availability}

Code and data to reproduce the results in this study will be made publicly available upon publication.

\section*{Acknowledgements}

This work was supported by the USDA Forest Service (FO-SFG-2673). The findings and conclusions in this publication are those of the author(s) and should not be construed to represent any official US Department of Agriculture or US Government determination or policy. Funding was also provided by NSF DEB-2213565 and NASA CMS grants Hayes (CMS 2020).

\bibliographystyle{apalike}
\bibliography{JZIG_ref}

\appendix

\section{Gibbs Sampler and Laplace Approximations}\label{app:GibbsNLaplace}

First we declare our notation.

\begin{enumerate}
    \item \textbf{Jointly sample $\btheta_y~|~\cdots$ where $\btheta_y = [\alpha_y~\beta_{y}~\bw_y]^T$ with a Laplace approximation.} \\ 
    
    Let $\tilde{\bA}_y = [\bone~\bA_x(\bB_y)\bw_x~\bA(\bs_y)]$ be the design matrix, where 
    \begin{equation}
        \log\left(\bmu_y(\bs_y|\btheta_y)\right) = \tilde{\bA}_y\btheta_y;~~\text{(see eqn. \ref{eq:muy})}
    \end{equation}, and $\tilde{\bQ}_y = \mathrm{blockdiag}(0, 0, \bQ_y)$ be the prior precision for $\btheta_y$. Further, define 
    \begin{equation}
        \bd_y(\btheta_y) = \by(\bs_y)\oslash\bmu_y(\bs_y|\btheta_y)
    \end{equation}
    as a $n_y\times 1$ vector where $\oslash$ represents element-wise division. Then we have the following gradient and Hessian of the conditional posterior
    \begin{align}
        \nabla p(\btheta_y | \cdots) &= \frac{1}{\phi_y}\tilde{\bA}_y^T(\bd_y(\btheta_y) - \bone) - \tilde{\bQ}_y\btheta_y\\
        \bH(\btheta_y) &= -\left(\tilde{\bQ}_y + \tilde{\bA}_y^T\mathrm{diag}(\bd_y)\tilde{\bA}_y\right).
    \end{align}
    Let $\btheta_y^{(0)}$ be some initial guess. The posterior mode, $\widehat{\btheta}_y$, is found via a Newton-Raphson procedure,
    \begin{equation}
        \btheta_y^{(j+1)} = \btheta_y^{(j)} - \bH(\btheta_y^{(j)})^{-1} \nabla p(\btheta_y^{(j)} | \cdots),
    \end{equation}
    until some convergence criterion is met. Then the Laplace approximation of the conditional is posterior is
    \begin{equation*}
        \btheta_y~|~\cdots \sim \mathrm{MVN}\left(\widehat{\btheta}_y,~-\bH(\widehat{\btheta}_y^{-1}) \right).
    \end{equation*}

    The above procedure can be accomplished with repeated Cholesky decompositions of the Hessian. Because the prior precision, $\tilde{\bQ}_y$, and design matrix, $\tilde{\bA}(\bs_y)$, are sparse, the Hessian is as well. Thus the decompositions can be performed efficiently with sparse matrix routines.
    \item \textbf{Jointly sample $\btheta_z~|~\cdots$ where $\btheta_z = [\alpha_z~\beta_{z}~\bw_z]^T$ with a Laplace approximation.} \\ 
    
    Let $\tilde{\bA}_z = [\bone~\bA_x(\bB_z)\bw_x~\bA(\bs_z)]$ be the design matrix, where 
    \begin{equation}
    \mathrm{logit}\left(\bmu_z(\bs_z|\btheta_z)\right) = \tilde{\bA}_z\btheta_z ; ~~ \text{(see eqn. \ref{eq:muz})}
    \end{equation}
    and $\tilde{\bQ}_z = \mathrm{blockdiag}(0, 0, \bQ_z)$ be the prior precision for $\btheta_z$. Further, define 
    \begin{equation}
        \bd_z(\btheta_z) = \bmu_z(\bs_z|\btheta_z) \odot \left(\bone - \bmu_z(\bs_z|\btheta_z)\right)
    \end{equation}
    be a $n_z\times 1$ vector where $\odot$ represents element-wise multiplication. Then we have the following gradient and Hessian of the conditional posterior
    \begin{align}
        \nabla p(\btheta_z | \cdots) &= \tilde{\bA}_z^T\left(\bz(\bs_z) - \bmu_z(\bs_z|\btheta_z) \right)\\
        \bH(\btheta_z) &= -\left(\tilde{\bQ}_z + \tilde{\bA}_z^T\mathrm{diag}\left(\bd_z(\btheta_z)\right)\tilde{\bA}_z\right).
    \end{align}
    Then the Laplace approximation follows the same format as Step 1 of the Gibbs sampler.
    \item \textbf{Jointly sample $\btheta_x~|~\cdots$ where $\btheta_x = [\alpha_x~\bw_x]^T$ with a Laplace approximation.} \\ 

    Analytically, this is the most difficult step, as $\bw_x$ appears in the likelihood of $\bx(\bs_x)$, $\by(\bs_y)$ and $\bz(\bs_z)$. Let
    \begin{align}
        \tilde{\bA}_x &= \left[\bone~~\bA_x(\bB_x)\right], \\
        \tilde{\bA}_{yx} &= \left[\bzero~~\beta_y\bA_x(\bB_y)\right], \\
        \tilde{\bA}_{zx} &= \left[\bzero~~\beta_z\bA_x(\bB_z)\right], 
    \end{align}
    be the design matrices, where
    \begin{align}
        \log\left(\bmu_x(\bB_x|\btheta_x)\right) &= \tilde{\bA}_x\btheta_x, \\
        \log\left(\bmu_y(\bs_y|\btheta_x)\right) &= \alpha_y\bone + \bA(\bs_y)\bw_y + \tilde{\bA}_{yx}\btheta_x, \\
        \mathrm{logit}\left(\bmu_z(\bs_z|\btheta_x)\right) &= \alpha_z\bone + \bA(\bs_z)\bw_z + \tilde{\bA}_{zx}\btheta_x.
    \end{align}
    Define the vectors
    \begin{align}
        \bd_x(\btheta_x) &= \bx(\bB_x) \oslash \mu_x(\bB_x|\btheta_x), \\
        \bd_y(\btheta_x) &= \by(\bs_y) \oslash \mu_y(\bs_y|\btheta_x), \\
        \bd_z(\btheta_x) &= \bmu_z(\bs_z|\btheta_x) \odot \left(\bone - \bmu_z(\bs_z|\btheta_x)\right).
    \end{align}

    Then the gradient and Hessian of the conditional posterior are
    \begin{align}
        \begin{split}
        \nabla p(\btheta_x~|\cdots) &= \frac{1}{\phi_t}\tilde{\bA}_x^T\left(\bd_x(\btheta_x) - \bone\right) + \frac{1}{\phi_y}\tilde{\bA}_{yx}^T\left(\bd_y(\btheta_x) - \bone\right) \\
        &\qquad + \tilde{\bA}_{zx}^T\left(\bz(\bs_z) -\bmu_z(\bs_z|\btheta_x)\right)
        \end{split} 
        \\[2ex]
        \begin{split}
        \bH(\btheta_x) &= \tilde{\bQ}_x + \tilde{\bA}_x^T\mathrm{diag}\left(\bd_x(\btheta_x)\right)\tilde{\bA}_x + \tilde{\bA}_{yx}^T\mathrm{diag}\left(\bd_y(\btheta_x)\right)\tilde{\bA}_{yx} \\
        &\qquad + \tilde{\bA}_{zx}^T\mathrm{diag}\left(\bd_z(\btheta_x)\right)\tilde{\bA}_{zx}
        \end{split}
    \end{align}

    Then the Laplace approximation follows the same procedure as Step 1 of the Gibbs sampler.    
    \item \textbf{Sample $\phi_y$ using a Metropolis-Hastings step}\\ 

    The conditional posterior for $\phi_y$ is
    \begin{equation}
        p(\phi_y~|~\cdots) \propto f\left(\by(\bs_y)~|~\bmu_y(\bs_y),~ \bphi_y\right)\cdot\frac{1}{\phi_y} ,
    \end{equation}
    where $f\left(\by(\bs_y)~|~\bmu_y(\bs),~ \bphi_y\right)$ is the density for $n_y$ iid Gamma variables with means $\bmu_y(\bs)$ and constant dispersion $\phi_y$. The factor $1/\phi_y$ is the posterior is due to the flat prior for $\log(\phi)$. Variable $\phi_y$ can be sampled with a Metropolis-Hastings step.
    \item \textbf{Jointly sample $\phi_x,~\phi_g$ using a Metropolis-Hastings step}\\ 

    The joint conditional posterior for $\phi_x,~\phi_g$ is
    \begin{equation}
        p(\phi_x,~\phi_g~|~\cdots) \propto f\left(\bx(\bB_x)~|~\bmu_x(\bB_x),~ \phi_x\bone + \frac{1}{\phi_g}\bN(\bB_x)\right)\cdot\frac{1}{\phi_x}\cdot\frac{1}{\phi_g} ,
    \end{equation}
    where $f\left(\bx(\bB_x)~|~\cdots\right)$ is the density for $n_x$ iid Gamma variables with means given by the vector $\bmu_x(\bB_x)$ and dispersions given by the vector $\phi_x\bone + \frac{1}{\phi_g}\bN(\bB_x)$. The variables can be jointly sampled with a Metropolis-Hastings step.
    \item \textbf{Sample the Mat\'ern parameters using Metropolis-Hastings steps}
    We give the procedure for jointly sampling $\sigma_x,~\rho_x~|~\cdots$. The process for the Mat\'ern parameters associated with $\bw_z$ and $\bw_y$ is parallel. The joint conditional posterior is
    \begin{equation}
        p(\sigma_x,~\rho_x~|~\cdots) \propto |\bQ_x(\sigma_x,\rho_x)|^\frac{1}{2}\exp\left(\frac{1}{2}\bw_x^T\bQ_x(\sigma_x,\rho_x)^{-1}\bw_x\right)\pi(\sigma_x, \rho_x),
    \end{equation}
    where $\pi(\sigma_x, \rho_x)$ is the joint penalized-complexity prior described in Section \ref{sec:inference}. The variables can be jointly sampled with a Metropolis-Hastings step.

\end{enumerate}

\end{document}